\documentclass[12pt]{article}
\usepackage{amsmath}
\usepackage{amsfonts}
\usepackage{amssymb}
\usepackage{upgreek}
\usepackage{makeidx}
\usepackage[utf8]{inputenc}
\usepackage[english]{babel}
\usepackage[square,numbers,sort&compress]{natbib}
\usepackage{graphicx}
\usepackage{lmodern}
\usepackage{color}
\usepackage{tabto}
\usepackage{bm}
\usepackage{float}
\usepackage{fancyhdr}
\usepackage{multirow}
\usepackage{braket}
\usepackage[a4paper, total={6.5in, 9.5in}]{geometry}
\usepackage{geometry}
\usepackage{setspace}
\usepackage{siunitx}
\usepackage{pdfpages}
\usepackage[inkscapelatex=false]{svg}
\usepackage[labelfont=bf]{caption}
\author{}
\usepackage{color}
\usepackage[hidelinks]{hyperref}
\numberwithin{equation}{section}
\usepackage{caption}
\usepackage{subcaption} 
\usepackage{authblk}
\usepackage{multicol}

\newcommand{\onlinecite}[1]{\hspace{-1 ex} \nocite{#1}\citenum{#1}} 

\title{\large \textbf{Sensing single-molecule magnets with nitrogen vacancy centers}}

\author[1,$^\dag$]{\small Ariel Smooha}
\author[1,$^\dag$]{Jitender Kumar}
\author[1]{Dan Yudilevich}
\author[1]{John W. Rosenberg}
\author[2]{Valentin Bayer}
\author[3]{Rainer Stöhr}
\author[3]{Andrej Denisenko}
\author[4]{Tatyana Bendikov}
\author[4]{Anna Kossoy}
\author[4]{Iddo Pinkas}
\author[5,$^\ddag$]{Hengxin Tan}
\author[5]{Binghai Yan}
\author[6,7]{Biprajit Sarkar}
\author[2]{Joris van Slageren}
\author[1,*]{Amit Finkler}

\affil[1]{\footnotesize Department of Chemical and Biological Physics, Weizmann Institute of Science, 7610001 Rehovot, Israel}
\affil[2]{Institute of Physical Chemistry, University of Stuttgart, Pfaffenwaldring 55, 70569 Stuttgart, Germany}
\affil[3]{Third Institute of Physics, IQST and
ZAQuant, University of Stuttgart, 70569 Stuttgart, Germany}
\affil[4]{Department of Chemical Research Support, Weizmann Institute of Science, Rehovot 7610001, Israel}
\affil[5]{Department of Condensed Matter Physics, Weizmann Institute of Science, Rehovot 7610001, Israel}
\affil[6]{Institute of Inorganic Chemistry, University of Stuttgart, Pfaffenwaldring 55, 70569 Stuttgart, Germany}
\affil[7]{Institut für Chemie und Biochemie, Freie Universität Berlin, 14195 Berlin, Germany}

\affil[$^\dag$]{These authors contributed equally to this work}
\affil[$^\ddag$]{\footnotesize{Present address: School of Physics and Astronomy, Shanghai Jiao Tong University, Shanghai, China}}
\affil[*]{\small amit.finkler@weizmann.ac.il}
\date{}

\begin{document}
\maketitle

\section*{Abstract}
Single-molecule magnets (SMMs) are molecules that can function as nanoscale magnets with potential use as magnetic memory bits. While SMMs can retain magnetization at low temperatures, characterizing them on surfaces and at room temperature remains challenging and requires specialized nanoscale techniques. Here, we use single nitrogen-vacancy (NV) centers in diamond as a highly sensitive, broadband magnetic field sensor to detect the magnetic noise of cobalt-based SMMs deposited on a diamond surface. We measure the NV relaxation and decoherence times at 296\,K and at 5-8\,K, observing a significant influence of the SMMs on them. From this, we infer the SMMs' magnetic noise spectral density (NSD) and underlying magnetic properties. Moreover, we observe the effect of an applied magnetic field on the SMMs’ NSD at low temperatures. The method provides nanoscale sensitivity for characterizing SMMs under realistic conditions relevant to their use as surface-bound memory units.
 
\vspace{1cm}
\textbf{\\Keywords}: Nitrogen vacancy centers, single-molecule magnets, quantum sensing, spin relaxometry, noise spectrum density

\clearpage

\def\arraystretch{1}

\newpage

\onehalfspacing 

\newpage

\renewcommand\theequation{\arabic{equation}}
\vspace{-0.3cm}
A single-molecule magnet (SMM) is a molecule that can behave as an individual nanomagnet. SMMs are promising candidates for magnetic data storage with ultra-high data densities due to their nanometer size. They consist of an inner core of one or more metal ions with a surrounding shell of organic ligands~\cite{Gatteschi2006} that can be tailored to bind them on surfaces~\cite{Bogani2008,Mannini2009}. 
Due to their mesoscopic size, they can be used to study the transition from the classical to the quantum mechanical regime, e.g., effects such as quantum tunneling of magnetization at low temperatures~\cite{Friedman2010}. 

Since the discovery of SMMs in 1993~\cite{Sessoli1993a, Sessoli1993}, these materials have attracted considerable interest for their potential applications in quantum computing~\cite{Stepanenko2008, Jenkins2013} and spintronics~\cite{Bogani2008, Hymas2019}.
However, a major challenge of utilizing SMMs for such applications is their fast spin dynamics at elevated temperatures due to stochastic magnetic fluctuations.
One of the key parameters to characterize SMMs is by their blocking temperature, $T_\text{B}$. Below this temperature, the magnetic moment of the molecule will be thermally stable (or `blocked'), and at higher temperatures, it behaves like a superparamagnet where the thermal fluctuations dominate, such that the average magnetization will be zero in the absence of magnetic field. 

Indeed, over the last three decades, there has been a constant effort to increase the blocking temperature of SMMs~\cite{Vieru2023}.
Here, another important aspect is the challenge in their detection and characterization at the nanoscale. Conventional methods for SMMs' characterization include superconducting quantum interference device (SQUID) magnetometry~\cite{Sessoli1993, Gatteschi2003}, electron spin resonance (ESR)~\cite{MorenoPineda2014}, inelastic neutron scattering (INS)~\cite{Garlatti2020}, and X-ray spectroscopy~\cite{Moroni2003}.
Other techniques include M\"ossbauer spectroscopy~\cite{Cini2018} and magnetic force microscopy (MFM)~\cite{Serri2017}. However, these techniques typically require either a macroscopic amount of material, low temperatures, or an ultra-high vacuum, and have a limited detection bandwidth for magnetic fluctuations. Furthermore, these techniques are less optimal for studying SMMs deposited on a surface, a geometry where they can practically function as memory units.

Here we demonstrate that it is possible to sense SMMs at nanoscale volumes using a a quantum sensor in the form of a single nitrogen vacancy center in diamond (NV), that allows us to measure the spectral density of magnetic noise of these molecules when applied on the diamond's surface. By comparing the nanoscale measurements to bulk ones \cite{Rechkemmer2016}, we show below the differences observed between them and provide a path to further exploration of other SMMs.

The NV center covers ten orders of magnitude of frequency bandwidth, ranging from sub-Hz up to the GHz regime, and functions at a broad range of temperatures~\cite{Eike2014-2}.
Moreover, the NV center is capable of sensing small magnetic moments outside of the diamond crystal, down to a single electron spin~\cite{Grotz2011, Grinolds2013, Shi2015}. It consists of a substitutional nitrogen and an adjacent vacancy, with a nanoscopic detection volume~\cite{Staudacher2013}. In its negatively charged state, it is a spin-1 system with spin-dependent photoluminescence (PL), enabling optically-detection of magnetic fields~\cite{Gruber1997, Doherty2013}. 

\begin{figure}[ht!]
\centering

\includegraphics[width=0.55\textwidth]{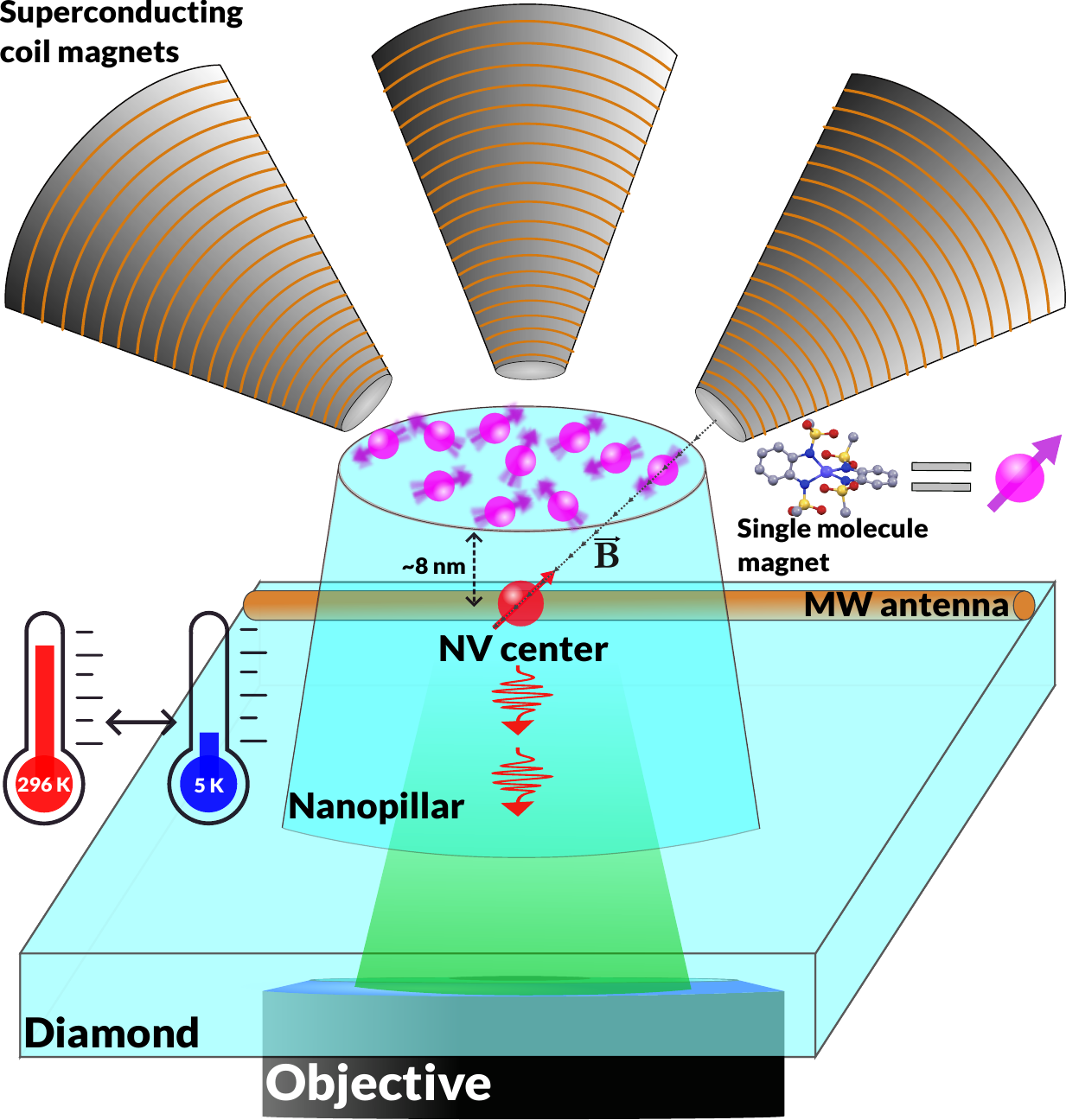}

\caption {\textbf{General scheme of the low-temperature NV setup (not to scale)}. Single-molecule magnets are deposited on top of a diamond nanopillar ($\sim$450\,nm diameter). A confocal microscope setup is used to excite the NV center with a 520\,nm laser, emitted photons are collected back, and measured. A DC magnetic field is applied with three superconducting coils. The LT system can reach 5\,K and resides in ultra-high vacuum (UHV) conditions. A microwave (MW) antenna is used for spin state manipulation.} 
\vspace{-0.4cm}
\label{setup_scheme}
\end{figure}

\textbf{NV-based relaxation measurements}. To investigate the magnetic noise generated by surface-deposited SMMs, we employ both $T_1$ longitudinal relaxometry and $T_2$ decoherence using shallow NV centers in diamond ($8\pm3$ nm from the surface, see Fig.\,\ref{setup_scheme}). Although the mean magnetic field $\braket{B}$ produced by an SMM spin bath may average to zero, magnetic field fluctuations generate a non-zero RMS field $\sqrt{\braket{B^2}}$ with a random phase. Such stochastic magnetic fields are inherently challenging to detect at the nanoscale. However, NV centers provide a powerful avenue for sensing them by monitoring their quantum spin relaxation dynamics~\cite{Schmid-Lorch2015, Steinert2013}.
After initializing the NV center into a well-defined spin state, interactions with its surrounding environment lead to spin relaxation. The relaxation is induced by intrinsic components from spin impurities and the vibrational lattice dynamics, and extrinsic components from the environment, such as nearby spin systems, to which the NV can be deliberately exposed.
Relaxation in spin systems may occur through two primary channels: decoherence $T_2$, which is sensitive to low-frequency fluctuations (kHz-MHz), and longitudinal relaxation $T_1$, which is sensitive to noise at a frequency near the NV center Larmor frequency ($\sim2.87$ GHz at low fields).
\\

\textbf{\\Sensing single-molecule magnets}. To sense cobalt-based SMMs, we deposited them onto a diamond membrane hosting shallow nitrogen-vacancy (NV) centers (see Sec.\,S8 in the SI for a detailed description). The full chemical formula of the SMM tetrahedral complex is $(\text{HNEt}_3)_2[\text{Co}^{\text{{II}}}(\text{L}^{2-})_2]$ where the ligand $L$ stands for 1,2-bis(methanesulfonamido)benzene~\cite{Rechkemmer2016}. Based on a 1 mM solution (in acetonitrile) and a spherical-cap geometry, the NV is expected to sense $\sim240$ molecules, within an effective sensing radius of 20 nm obtained from a simulation (sensing volume of $\sim$(20\,nm)$^3$, see Sec.\,S1 in the SI). Similar NV sensing ranges were also experimentally found when sensing transition metals \cite{Flinn2023}.

\begin{figure}[ht!]
\centering

\includegraphics[width=0.70\textwidth]{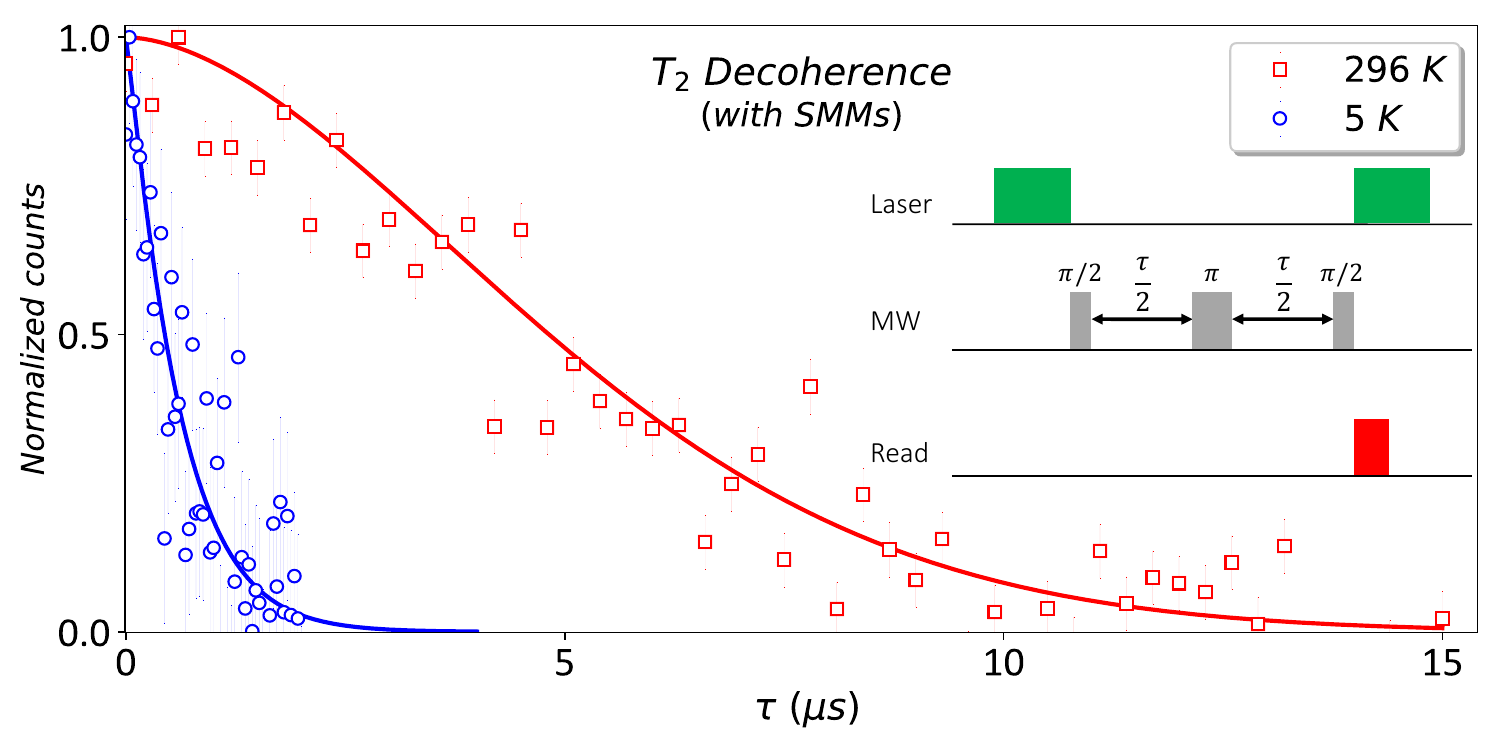}

\caption{ \textbf{The effect of the cobalt-based SMMs at RT and 5\,K}. The NV center (NV17) $T_2$ coherence curve, in the presence of SMMs, at 296\,K (red) and 5\,K (blue), with $T_2=5.9\pm0.3~\text{\textmu s}$ and $T_2=0.62\pm0.09~\text{\textmu s}$, respectively. The data were fitted based on Sec.\,S3 in the SI. Inset: the $T_2$ pulse sequence.}

\label{T2_NV_17}
\end{figure}

We thoroughly characterize the deposited cobalt-based SMM layer on diamond with three spectroscopic techniques. Specifically, X-ray diffraction (XRD) and Raman spectroscopy indicate that the deposited layer  retains the SMM molecular structure, and X-ray photoelectron spectroscopy (XPS) demonstrated that the cobalt is found in the Co$^{2+}$ oxidation state (see Sec.\,S2 and Sec.\,10 in the SI).
A confocal microscopy setup was used to optically polarize and read out individual NV centers. We first carried out $T_2$ coherence time measurements at room temperature (RT, $\sim$296\,K) and low temperature (LT, $\sim$5\,K) in the presence of SMMs. Measurements were repeated across multiple NV centers to ensure reproducibility and to account for variations in local environments. In a $T_2$ measurement, we utilize the spin echo pulse sequence for tracking the decay profile of the superposition state $\ket{\psi}=\frac{1}{\sqrt{2}}\left( \ket{0}+\ket{1} \right)$ with time. A representative $T_2$ measurement is shown in Fig.\,\ref{T2_NV_17}. In the presence of SMMs, we observed a significant reduction by approximately one order of magnitude in the coherence time $T_2$ of the NV center when cooling from 296\,K (RT) to 5\,K (LT), under a low magnetic field ($B_0\sim20$ G). 

We also performed $T_1$ measurements at RT and 5\,K by optically initializing the NV center to the $\ket{0}$ state and tracking the relaxation profile of the system without MW radiation\,\cite{Tetienne2013a}. While the $T_2$ measurements exhibited a clear and pronounced temperature dependence, the $T_1$ measurements showed only a modest increase in relaxation time at 5\,K compared to RT (Fig.\,S4 in the SI). 

\begin{figure}[ht!]
     \centering
     \begin{subfigure}[b]{0.49\textwidth}
         \centering
         \includegraphics[width=1\textwidth]{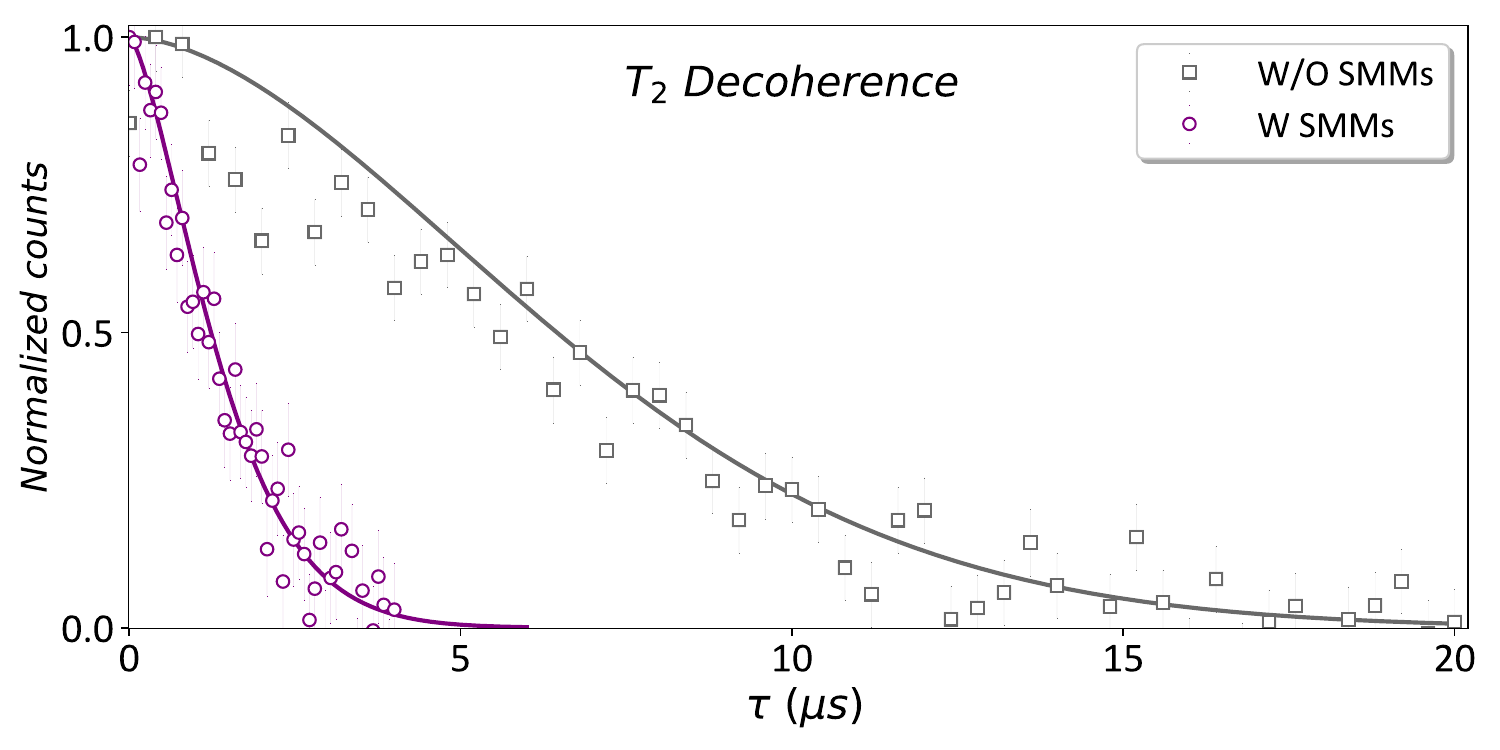}
         \vspace{-0.7cm}
         \caption{}
         \label{T2_LT_wo_NV_24}
     \end{subfigure}

    \begin{subfigure}[b]{0.49\textwidth}
         \centering
    	 \includegraphics[width=1\textwidth]{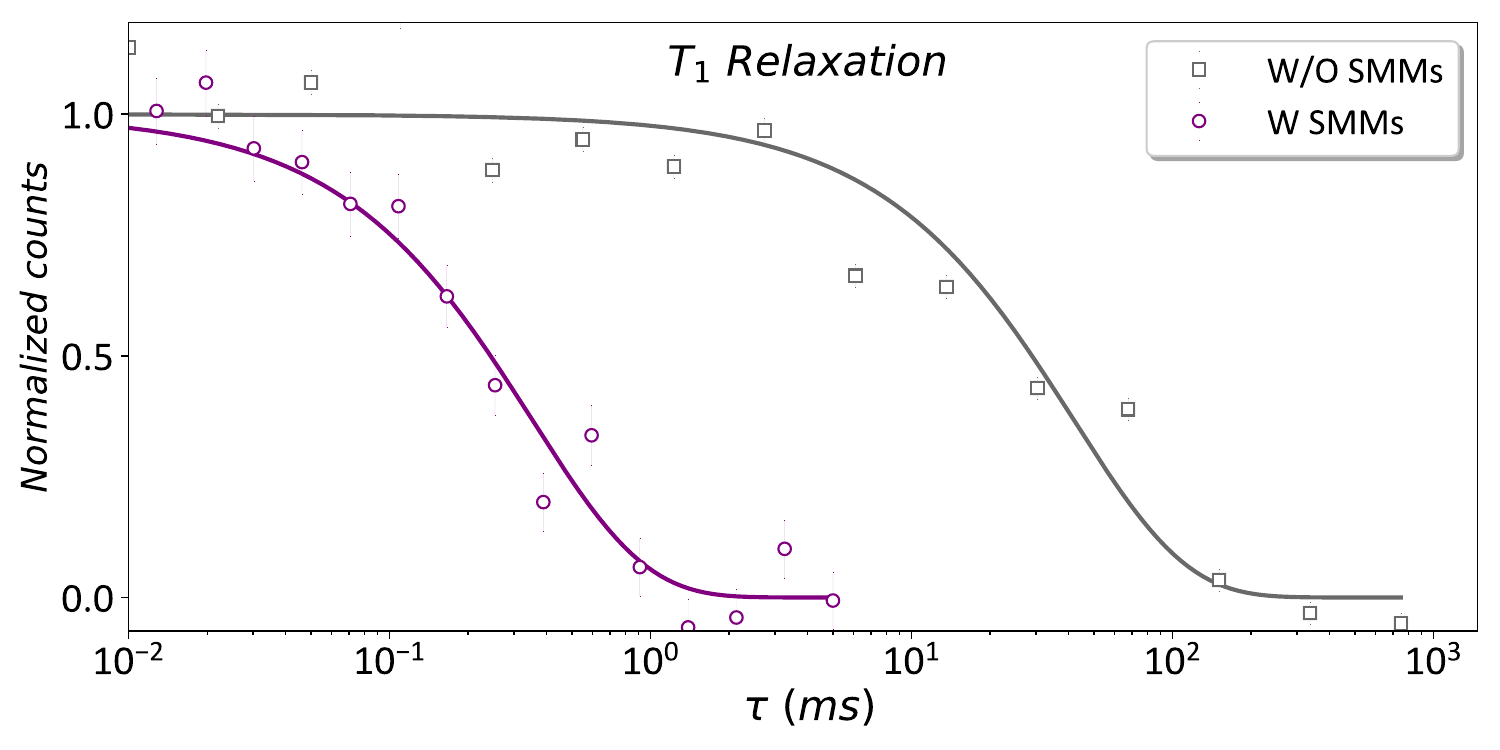}
         \vspace{-0.7cm}
         \caption{}
         \label{T1_LT_wo_NV_24}
     \end{subfigure}
        \caption{ \textbf{The effect of the cobalt-based SMMs at $\sim$5\,K}. 
        \textbf{(a)} The NV center (NV24) $T_2$ coherence curves in the presence (purple) and absence (gray) of SMMs, with $T_2=0.80\pm0.04\,\text{\textmu s}$ and $T_2=4.0\pm0.2\,\text{\textmu s}$, respectively.
        \textbf{(b)} $T_1$ relaxation curves (NV24) in the presence (purple) and absence (gray) of SMMs with $T_1=0.35\pm0.05$\,ms and $T_1=22\pm4$\,ms, respectively. The data were fitted based on Sec.\,S3 in the SI.
        }
        \label{SMMs_effect_LT}
\end{figure}

To confirm that the observed effects at 5\,K were induced by the SMMs and not by other possible sources of magnetic noise, such as surface or lattice spin species, we conducted comparative $T_2$ measurements at 5\,K after removing the SMMs from the diamond surface via an acid-cleaning protocol~\cite{Sangtawesin2019}. A representative post-cleaning $T_2$ measurement is presented in Fig.\,\ref{T2_LT_wo_NV_24}. In the absence of SMMs, we observed a substantial recovery of the $T_2$ coherence times at 5\,K, with values increasing by up to a factor of five compared to those measured in the presence of SMMs. 
We also conducted comparative $T_1$ measurements at 5\,K after removing the SMMs. In this case, we observed a pronounced reduction in the $T_1$ relaxation time of the NV centers at 5\,K in the presence of SMMs, amounting to approximately one to two orders of magnitude relative to the cleaned sample (Fig.\,\ref{T1_LT_wo_NV_24}).
These findings suggest that the enhanced relaxation rate observed at 5\,K arises from magnetic field fluctuations originating from the SMMs.  The overall trend is consistent across all measured NVs (see Sec.\,S4 in the SI).

The observed trend can be explained based on a general model for electronic spin noise, describing the behavior of magnetic fluctuators. 
In this model, the dynamics of the magnetic spin noise can be described as an Ornstein-Uhlenbeck process~\cite{deLange2010,Dobrovitski2009}, which is a stationary Gauss-Markov process, with the following autocorrelation function 
\begin{equation}
A(t)=\left\langle B(0)B(t)\right\rangle =\left\langle B^{2}\right\rangle \exp\left(-\frac{t}{\tau_{c}}\right)
\label{autocorrelation_}
\end{equation}

\noindent where $B(t)$ is the time-dependent magnetic field induced by the noise bath, and $\tau _c$ is the correlation time, which is the `memory time' of the environmental noise. The Fourier transform of Eq.\,\ref{autocorrelation_} yields the normalized magnetic noise spectrum density (NSD)~\cite{Schmid-Lorch2015} 
\begin{equation}
S \left(\omega,\ T,\ E_{a}\right)=\frac{2}{\pi}\frac{\tau_c\left(T,\ E_{a}\right)}{1+\tau_c\left(T,\ E_{a}\right)^{2}\omega^{2}}
\label{noise_spectrum}
\end{equation}
where we introduce the anisotropy energy barrier of a single-molecule magnet $E_a$. 
In the presented case of the cobalt-based SMM, the relaxation rate is governed by Raman and Orbach processes as follows~\cite{Rechkemmer2016}
\begin{equation}
\tau_c^{-1}=\underset{\text{Raman}}{CT^{n}}+\underset{\text{Orbach}}{\tau_{0}^{-1}\exp\left(-E_{a}/k_{B}T\right)} .
\label{raman_orbach}
\end{equation}
Taking the Raman process into account is essential for an accurate fit of the model, as was shown in a previous study on these SMMs by Rechkemmer et al.~\cite{Rechkemmer2016} on pressed powder pellets. In general, for different magnetic species, the Orbach process is sufficient for modeling the system~\cite{Eike2014-2,Schmid-Lorch2015}. However, in our case, the behavior at low temperature deviates significantly from an Orbach process, suggesting that a Raman process plays a significant role at low temperature.
According to Ref.\,\onlinecite{Rechkemmer2016}, the parameters that fit Eq.\,\ref{raman_orbach} are $C=0.088\pm 0.009$, $n=3.65\pm0.04$, and $\tau^{-1}_0=(9.1\pm 0.6)\cdot 10^9~\text{s}^{-1}$, where the last is known as the inverse attempt frequency. The energy barrier, which was fixed and verified by independent infrared spectroscopy, is $E_a = 230~ \text{cm~}^{-1}=28.52\text{~meV}$~\cite{Rechkemmer2016}. The blocking temperature of the SMM in its bulk form has not been reported. However, Rechkemmer et al. \cite{Rechkemmer2016} observed magnetic hysteresis at 1.8\,K, indicating that the blocking temperature is likely close to this value. 

The effect of the SMMs' noisy spin bath on the relaxation rate of the NV center is incorporated through the coherence signal $e^{-\chi(t)}$ where $\chi(t)$ is known as the coherence function~\cite{Biercuk2011}. This function is dependent on the filter function $F(\omega)$, which in turn is determined by the pulse sequence, and the NSD of the surrounding spin bath $S(\omega)$ as (derivation in Sec.\,S5 in the SI)~\cite{Sousa2009, Davis2023} 
\begin{equation}
\chi(t)=\int{d\omega}S(\omega)F(\omega)
\label{chi}
\end{equation}
such that we expect to have a faster relaxation rate as the overlap between the filter function $F(\omega)$ and the NSD $S(\omega)$ becomes larger.\\
In order to explain the influence of the SMMs on $T_i~(i=1,2)$ of the NV center, we use the following equation for the relaxation time dependence on the NSD\,\cite{Eike2014-2}:
\begin{equation}
\frac{1}{T_{i}}=\left(\frac{1}{T_{i}}\right)_{\text{int}}+\gamma_\text{NV}^{2} \left\langle B^{2}\right\rangle \int  S\left(\omega,\ T,\ E_{a}\right)F_{i}\left(\omega\right)d\omega,
\label{relaxation_rate}
\end{equation}
where $\gamma_{NV}$ is the NV electron spin gyromagnetic ratio, $\sqrt{\left\langle B^{2}\right\rangle}$ is the effective magnetic field at the NV position generated by the SMMs, and $F_{i}\left(\omega\right)$ is the filter function determined by the pulse sequence. In our case, the $T_1$ and $T_2$ based protocols are used for the characterization of the NSD behavior at the GHz and MHz regimes, respectively.

\begin{figure}[ht!]
     \centering
     \includegraphics[width=0.65\textwidth]{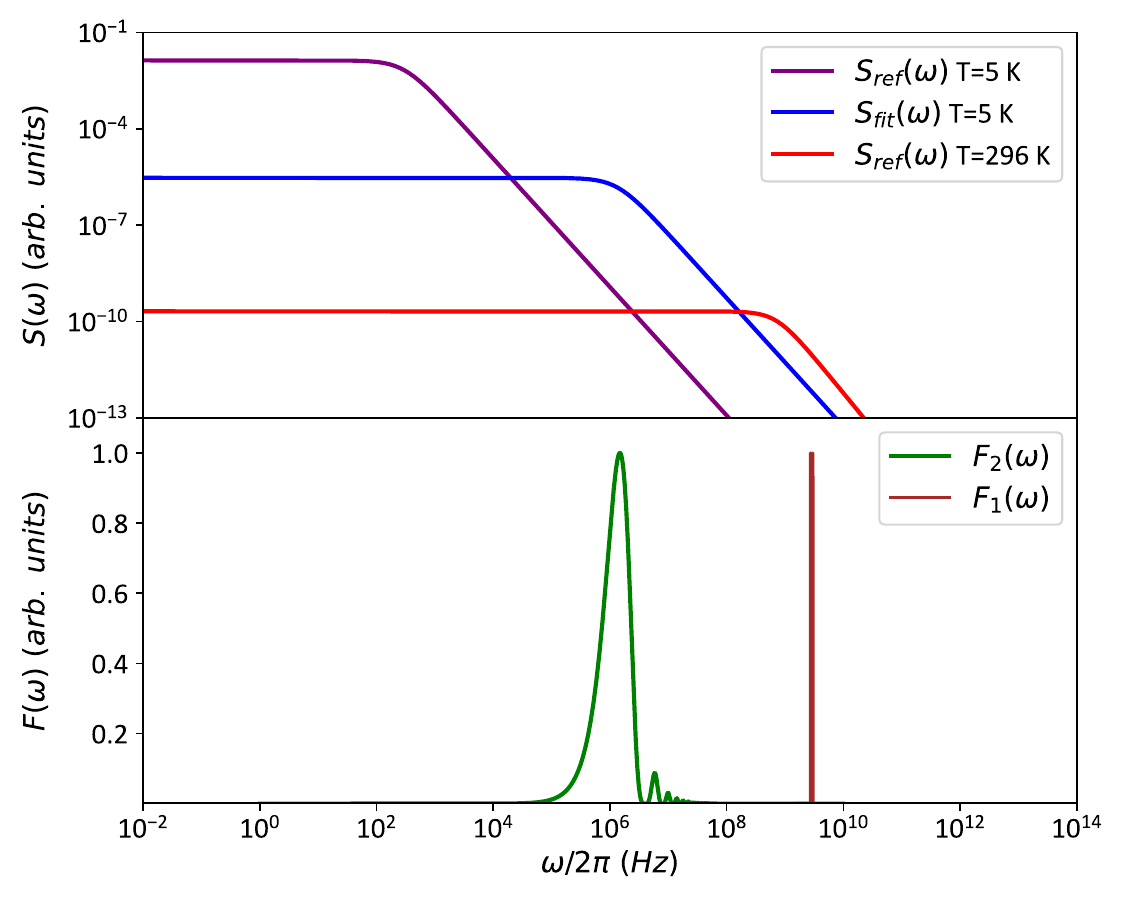}

     \caption{ \textbf{NSDs and filter functions}. The upper part depicts the normalized NSD of the cobalt-based SMM at 5\,K (purple) and 296\,K (red) based on reference data~\cite{Rechkemmer2016}. In blue is the fitted NSD at 5\,K (for NV22, based on Eq.\,S3 and Eq.\,S4 in the SI, Sec.\,S6). The lower part depicts the $T_2$ (green) and the $T_1$ (brown) filter functions.}
     \label{NSD_filters}
     \vspace{-0.1cm}
\end{figure}

In a $T_2$ measurement the filter function is~\cite{Schmid-Lorch2015}:

\begin{equation*}
F_{2}(\omega)=\frac{32}{\tau}\frac{\sin^{4}\left(\frac{\omega\tau}{4}\right)}{\omega^{2}},
\end{equation*}

\noindent with $\tau/2$ the interpulse delay, as shown in Fig.\,\ref{NSD_filters} (green curve). According to Eqs.\,\ref{noise_spectrum} and \ref{raman_orbach}, and using the fit parameters reported in Ref.\,\onlinecite{Rechkemmer2016}, the SMMs’ NSD exhibits a cutoff frequency above the GHz range at RT (Fig.\,\ref{NSD_filters}, red curve). At 5\,K, this cutoff shifts to lower frequencies, accompanied by an increase in noise amplitude (purple curve). 

We extract the correlation time of the Raman process by fitting the $T_1$ and $T_2$ experimental results at 5\,K of five NVs in the presence of SMMs. We obtain an average correlation time of $\tau_c=5\pm1$ \textmu s for these five NVs (see Sec.\,S6 in the SI for further details). This value is different than a previously published value of $\tau_c =21\pm2$ ms~\cite{Rechkemmer2016}, such that the Raman coefficient $C$ and power $n$ are different in this case.

This result is reasonable due to the different phonon spectra in our measurements. The cobalt-based SMMs, in our case, are not in the bulk state but diluted on a diamond surface (Sec.\,S2 in the SI). The amorphous nature of the drop-cast sample probably leads to distortions, destroying the highly axial nature of the anisotropy, leading to fast underbarrier processes.

Thus, since the Raman process typically involves molecular vibrations in such systems, we obtained a different relaxation rate than that for a bulk sample~\cite{Rechkemmer2016}. Moreover, varying values of the Raman power $n$, which correspond to differences in relaxation rates, have been previously reported in the literature~\cite{Lunghi2023, Novikov2015, Boulon2012}. 

In our case, we have a larger overlap at low temperatures between the filter function and the NSD in the frequency domain, as can be seen in Fig.\,\ref{NSD_filters} (blue curve). This accounts for the significant, nearly one order of magnitude reduction in $T_2$ observed in the presence of SMMs. The reduction in $T_2$ is in contrast to the opposite effect previously observed in bare diamonds, namely, an increase or no change in $T_2$ when lowering the temperature~\cite{Takahashi2008}, since thermal fluctuations of the spin bath decrease with temperature.

In a $T_1$ measurement, the filter function is given as~\cite{Schmid-Lorch2015}:

\begin{equation*}
F_{1}(\omega)=\sum_{i=\pm1}\frac{4\pi/T_{2}^{*}}{(2\pi/T_{2}^{*})^{2}+(\omega-\omega_{i})^{2}}
\end{equation*}
where $T_2^*$ is the dephasing time of the NV center ($\sim$2 \textmu s) and $\omega_i$ is the resonance frequency of the NV ($\sim2\pi\times$2.87 GHz at a low magnetic field) as shown in Fig.\,\ref{NSD_filters} (two sharp, closely spaced brown peaks that appear as a single peak). As a consequence, the spin probe can be sensitive to magnetic field fluctuations in the GHz regime.

On the one hand, the dominant $T_1$ relaxation mechanism of NV centers at RT involves two-phonon Orbach and Raman processes~\cite{Jarmola2012}. On the other hand, at 5\,K, the relaxation is temperature-independent and governed by cross-relaxation with neighboring spins. Hence, with our system on bare NVs, we expect to have longer $T_1$ values at RT than at 5\,K as also shown in the control measurements of Sec.\,S4 in the SI. 
However, we observed that in the presence of SMMs, when moving from RT to 5\,K, we do not obtain a significant difference in the $T_1$ values, and they remain similar to the RT values of a few hundred microseconds (Fig.\,S4b in the SI).  Thus, it implies that the SMMs have a major influence on $T_1$ at 5\,K such that relaxation processes induced by the SMMs are significant. This suggests an overlap between the $T_1$ filter function and the SMMs’ NSD components at both 5 K and RT, as can be seen in Fig.\,\ref{NSD_filters}. 

\begin{figure}[ht!]
     \centering
               \begin{subfigure}[b]{0.49\textwidth}
         \centering
         \includegraphics[width=1\textwidth]{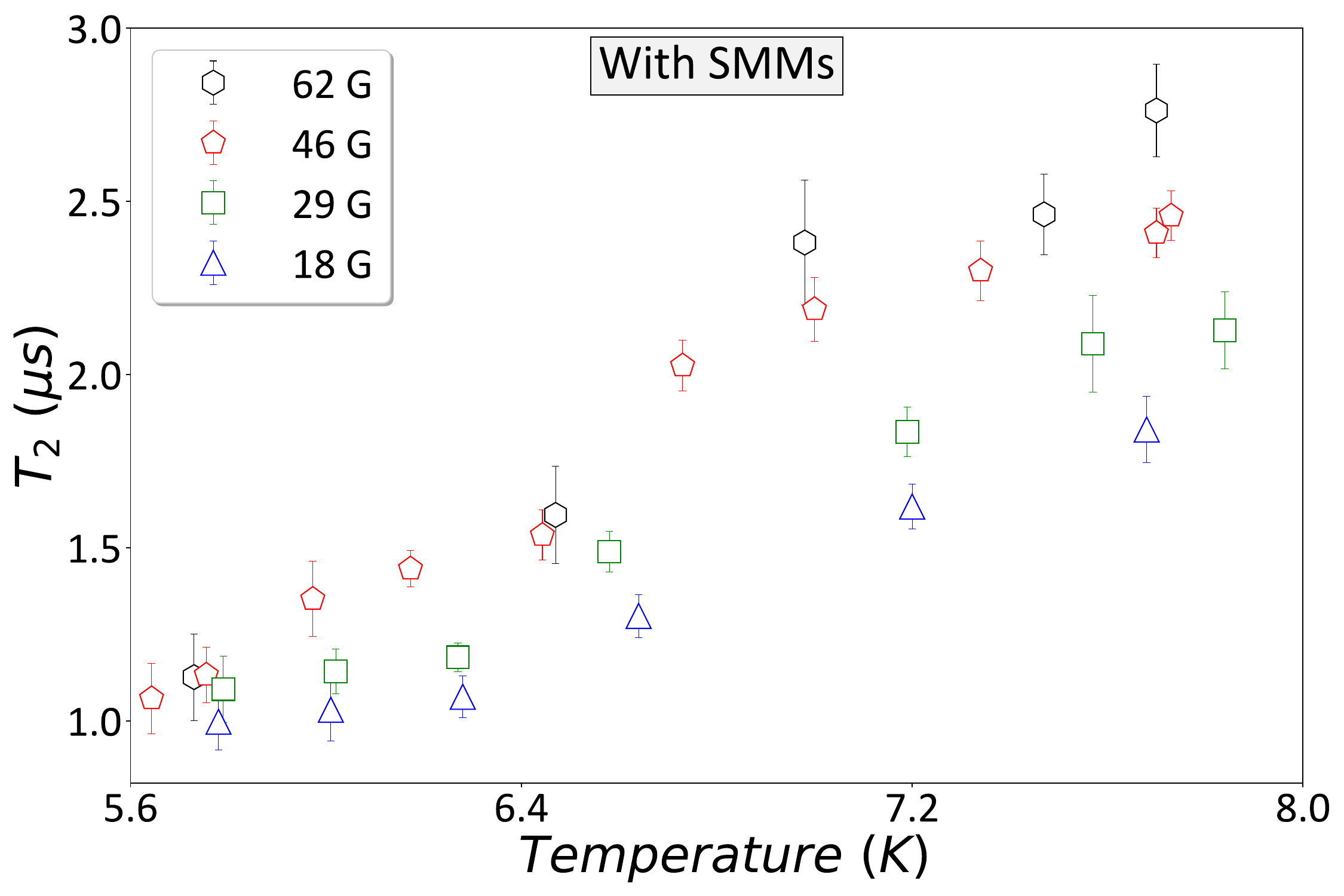}

         \caption{}
         \label{field_vs_current_smms}
     \end{subfigure}
     \hfill
     \hspace{0cm}
     \begin{subfigure}[b]{0.49\textwidth}
         \centering
         \includegraphics[width=1\textwidth]{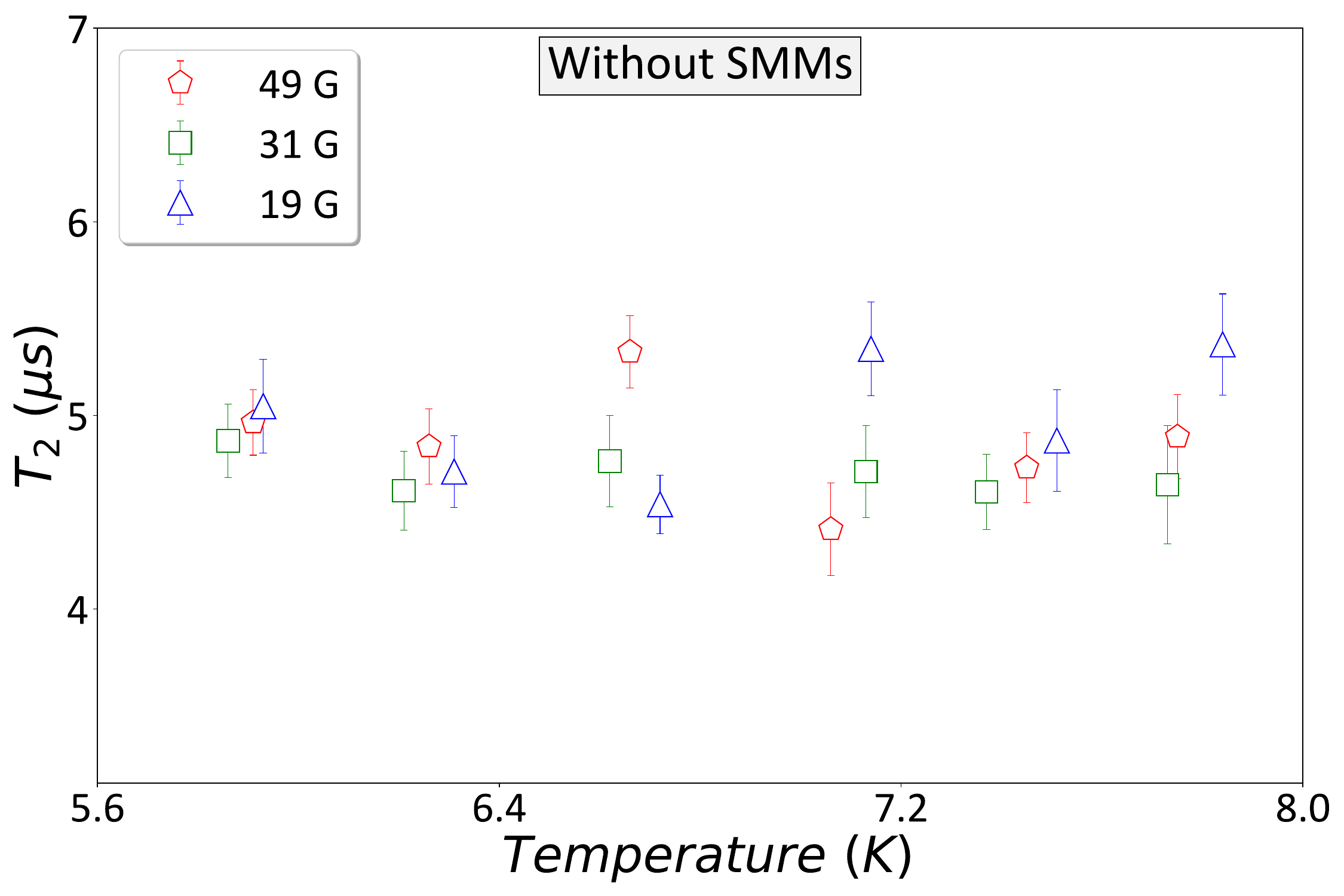}

         \caption{}
         \label{field_vs_current_reference}
     \end{subfigure}

        \caption{ \textbf{NV coherence $T_2$ vs. temperature at different magnetic fields.} Observing the  effect of applied magnetic fields on the spin dynamics of cobalt-based SMMs. 
        \textbf{(a)} Data with SMMs. $T_2$ values of NV22 as a function of temperature and magnetic field in the presence of SMMs. 
        \textbf{(b)} Reference data, without SMMs. $T_2$ values of NV22 as a function of temperature and magnetic field in the absence of SMMs. All data were fitted based on Sec.\,S3 in the SI. 
        }
        \label{T2_vs_temp_field}
\end{figure}

\textbf{\\Magnetic field dependence}.
We performed magnetic field- and temperature- dependent measurements with SMMs to investigate their effect on the NV center, providing further insight into the SMMs behavior. 
We observed a field-dependent impact of SMMs on the NV center. As shown in Fig.\,\ref{field_vs_current_smms}, we measured the $T_2$ values at different magnetic fields ranging from 18\,G up to 62\,G in a temperature window ranging from $\sim$8\,K to the base temperature of the experimental setup (Sec.\,S7 in the SI). 

Comparing the results, we can observe a significant increase in the coherence time $T_2$ as the magnetic field increases and also as the temperature increases.

At around 5.6\,K, there is almost no change in the coherence times at the different magnetic fields, but as the temperature increases, this change becomes observable. Comparing the $T_2$ values at about 7.7\,K at 18\,G (blue) to give $T_2=1.8\pm 0.1~$\textmu s  and 62\,G (black) to give $T_2=2.8\pm 0.1~$\textmu s, we have an increase of about 60\%. 

The monotonic increase in $T_2$ with rising temperature might be attributed to the reduction in the time-averaged magnetic moments of noise-inducing entities as the temperature moves further above the blocking temperature of the SMMs. Notably, strong temperature dependence in both frequency and magnitude dispersion has been observed in these cobalt-based SMMs during frequency- and temperature-dependent AC magnetic susceptibility measurements\,\cite{Rechkemmer2016}.

Reference measurements without SMMs (Fig.\,\ref{field_vs_current_reference}) suggest that this effect can be attributed to the SMMs.
First, the $T_2$ values are notably longer, consistent with our previous observations. Second, the $T_2$ values are constant over these temperature and magnetic field ranges, as expected for a bare NV center.

Thus, it strengthens the fact that the above-mentioned behavior stems from the presence of SMMs. 

In conclusion, we presented the sensing of cobalt-based single-molecule magnets using single NV centers. We characterize the magnetic noise spectrum of these molecules at RT and 5\,K and under varying magnetic fields. We observed a significant reduction in the coherence time $T_2$ and longitudinal relaxation time $T_1$ of the NV center in the presence of SMMs. We modeled the magnetic noise spectrum of the SMMs as a Gauss-Markov process, and used the coherence function to relate the NSD and the filter function to the relaxation profile of the NV center. With this, we could explain the effect of the SMM spin bath on the $T_1$ and $T_2$ relaxation times of the NV centers at different temperatures. Moreover, we were able to extract the correlation time of the SMMs bath at 5\,K. By acquiring an additional $T_1$ and $T_2$ dataset at another low temperature, it is further possible to determine the Raman coefficient and exponent. In this case, it may be better to work in the diluted regime, where dipolar interactions between the SMMs are less pronounced, since otherwise they add another contribution to the extracted correlation time. Our approach represents a novel methodology for extracting the parameters governing the Raman relaxation process of surface-deposited SMMs, which has not been reported in previous studies. 

We have also observed a significant variation in the $T_2$ of an NV center as a function of temperature in the presence of SMMs, particularly near the blocking temperature regime. In addition, a strong positive influence of an applied magnetic field on $T_2$ has been observed in this temperature regime. This indicates that applying a DC magnetic field modulates the noise profile of the SMMs, a noteworthy observation with potential implications for their use in storage technology. Going to even lower temperatures, i.e., below the blocking temperature, would allow not only to identify the structural form of the molecules next to the NV on the surface of the diamond, i.e., whether it is a molecular crystal or isolated molecules, but also to isolate the contribution of SMMs' intermolecular dipolar coupling, which is currently difficult to establish.

The method we presented here can help in the research and development of SMMs since we can sense them at the surface, at different temperatures, and at nanoscopic volumes. Moreover, this method can also be applied to the detection and characterization of other types of SMMs, as well as potential molecular qubits.

\normalsize

\section*{Acknowledgements}
We thank M.\,W.\,Doherty, O.\,Tal, J.\,S.\,Miller and L.\,Kronik for insightful discussions. We also thank I.\,Zohar and L.\,Schein-Lubomirsky for experimental assistance and proofreading the manuscript. IP is the incumbent of the Sharon Zuckerman research fellow chair.
AF acknowledges financial support from the Minerva Stiftung (Grant 714131) and the Israel Science Foundation (Grants 418/20, 419/20). AF also acknowledges support by the Kimmel Institute for Magnetic Resonance. AF is the incumbent of the Elaine Blond Career Development Chair in Perpetuity. JvS acknowledges funding from the German Science Foundation
(DFG, SL104/10-1, SA1840/9-1). This research is made possible in part by the historic generosity of the Harold Perlman Family. 

\subsection*{Data Availability}
The data that support the findings of this study are openly available at the following URL/DOI: \href{https://doi.org/10.34933/400b149f-da31-441b-aace-0e4b037205b8}{10.34933/400b149f-da31-441b-aace-0e4b037205b8}

\subsection*{Supporting Information}
The Supporting Information is available free of charge at the publisher's website (TBD).\\\\
Contents: Simulation of NV effective sensing radius and volume (S1); Comparison between bulk and drop cast SMMs (S2); $T_1$ and $T_2$ curves fitting (S3); Additional statistics and control measurements (S4); Derivation of the coherence function in the frequency domain (S5); Model fitting to the data (S6); Temperature control at LT (S7); Sample preparation (S8); SMMs overview (S9); XPS analysis (S10); Experimental setup (S11); \text{ab-initio} calculations (S12); The T\texorpdfstring{$_1$}{TEXT} protocol (S13); Quantum sensing with NV centers (S14).    

\onecolumn
\clearpage
\renewcommand{\refname}{\large References}
\bibliography{main_v2.bib}

\end{document}